\documentstyle[aps,twocolumn,floats,epsf]{revtex} 
\begin{document}  
\draft
\newcommand{\be}{\begin{equation}}
\newcommand{\ee}{\end{equation}}
\newcommand{\bra}{\langle}
\newcommand{\ket}{\rangle}
\newcommand{\del}{\partial}
\renewcommand{\arraystretch}{1.8}
\title{Excitation of Small Quantum Systems by High--Frequency Fields}
\author{Naama Brenner$^a$ and Shmuel Fishman$^{a,b}$}
\address{$^a$ Dept. of Physics, Technion, Haifa 32000, Israel\\
$^b$ Institute for Theoretical Physics, Universtiy of Califronia,
Santa Barbara, Ca. 93106, USA}  
\maketitle     
\begin{abstract}
The excitation by a high frequency field of
multi--level quantum systems
with a slowly varying density of states is investigated. 
A general approach to study such systems is presented.
The Floquet 
eigenstates are characterized on several energy scales. On a small 
scale, sharp universal quasi--resonances are found, whose shape is
independent of the field parameters and the details of the system.
On a larger scale an effective tight--binding equation is constructed 
for the amplitudes of these quasi--resonances. This equation is 
non--universal; two classes of examples are discussed in detail. 
\end{abstract}     
\par                                                                 

\vspace{0.5cm}  
The excitation of quantum systems
by external fields, is a fundamental problem in quantum mechanics.
Any interaction between matter and electromagnetic
fields is described, on the simplest level, by such a model.
Given the system at some initial condition, one would like to characterize
its energy absorption from the external field as a function of time.
In some cases, simple quantum systems can be described effectively by one
degree of freedom.
One--dimensional (1D) driven systems have been studied widely within 
the field of ``quantum chaos'' \cite{Varenna91,Haake91},
since they provide some of the simplest examples of quantum systems
which are chaotic in the classical limit. The phonomenon of 
dynamical localization,
where the classical energy absorption associated 
with chaotic motion is suppressed
by quantum interference effects \cite{Casati79},
has been of special interest
(For reveiew see S. Fishman, in \cite{Varenna91}).

In this work, a general approach for systems  
with a slowly--varying density of states, driven by a high--frequency
smooth periodic field, is presented.
We study the quasi--energy (Floquet) eigenstates, which are the stationary
states of the time dependent system 
\cite{Zeldovich67}, in the unperturbed representation,
and characterize their properties over several energy scales.
On the smallest scale, we find that, under
quite general conditions, these states are
composed of a ladder of sharp peaks, or ``quasi--resonances''
(QR). These are related to quantum nearly resonant transitions
between the energies of the undriven
system and should be distinguished from the classical resonances.
While the existence of
such peaks is well known \cite{Leopold88}, here they are derived from
a local exact solution, which enables the quantitative
description of their location and shape.
Surprisingly, the QR's turn out to have a universal
shape, independent of field parameters and of the details
of the system. In particular, their width is independent
of the driving field strength, a result which contrasts 
simple perturbation theory or a two--level approximation.
On a larger energy scale, our local solution
allows us to construct an effective equation for the envelope of amplitudes
superimposed on the QR's \cite{Oliveira94}.
The form of the local solution is used explicitly to obtain
a tight--binding equation on the lattice of QR peaks.
The parameters of this
equation are non--universal and are related to the
spectrum of the unperturbed
system and to the matrix elements of the perturbation. 
  
Investigation of the Floquet operator (the evolution
operator for one period), is very 
convenient for time periodic systems.
Its spectrum determines
the quantum dynamics on all time scales, similar to
the Hamiltonian for autonomous systems.
Although finding this spectrum
exactly is as difficult as solving the time--dependent
Schroedinger equation, here we characterize the eigenstates by
general qualitative properties, from
which properties of the dynamics follow.
This idea is inspired by 
the kicked rotor analogy \cite{kr1}, where a robust qualitative
feature of the quasi--energy states - namely their
exponential localization in energy space - explains the
quantum saturation in energy absorption. A formal
mapping was constructed to a 1D Anderson model,
which relies on special properties
of the kicked rotor; other driven systems require in general
a different treatment.

For systems with a slowly varying density of states,
a constant energy spacing is well defined on a local energy
scale, and the deviations of the spectrum
from harmonicity appear as an
adiabatic change of this spacing with energy.
This ``adiabatic nonlinearity'' of the spectrum is characteristic
of many 1D systems, of which several examples
are the hydrogen--like atom (\cite{Blumel90}, 
D. L. Shyepelyansky in \cite{Varenna91}),
charge bubbles in liquid helium 
\cite{Fetter76,Shimshoni87,Benvenuto91,Oliveira94} and 
surface electrons in a 2D metal in a perpendicular
magnetic field \cite{Prange68}.
This work presents a general framework for this class of
systems, from which some new results emerge and some known
results for special cases are confirmed.
We focus on 
a monochromatic driving, which is of high--frequency
compared to the typical frequencies of the system.
Thus, many unperturbed levels participate in the excitation
(it is important, however, that the spectrum is
discrete for our solution to hold).
The basic idea employed in this work, is to
solve the problem in a limited energy regime, where the
locally defined energy spacing can be considered constant. This local
solution relies on the exact quantum mechanical solution of an
integrable model \cite{lin1}. We then account for the large energy scales by 
exploiting the adiabatic dependence of parameters on energy.
A similar idea was previously applied to
the ``bubble'' model \cite{Brenner95}. 
Here a wider class of systems is studied 
(more details in \cite{Brenner96}).

Consider the following 1D
Hamiltonian in action--angle variables
$(I,\theta)$ of the bound system:
\be
{\cal H}~~=~~{\cal H}_0(I)~+~
k~V(I)g(\theta)\cos(\Omega t).  \label{hamil1}
\ee
\noindent It is assumed that ${\cal H}_0(I)$ and $V(I)$ are smooth
functions of $I$, and that $g(\theta)$ has a Fourier expansion with
smooth, slowly decaying components $G(m)$.
Our object is to calculate the
Floquet operator, and to characterize
its eigenstates in the unperturbed representation $|n\ket$, where
$I=n\hbar$. In the high--frequency regime,
the field period is a relatively short time scale and thus the semiclassical
approximation is expected to be very accurate.
Using also a leading order approximation to the classical trajectories
\cite{Percival70},
the Floquet eigenvalue equation is 
\be
e^{-i \frac{T}{\hbar}{\cal H}_0(n\hbar)}~
e^{-\frac{i}{\hbar}A(I;\hat{\theta})}|\psi_{\lambda}\ket~~=~~
e^{-i\lambda T}|\psi_{\lambda}\ket,
\label{kick}
\ee
where $T=2\pi/\Omega$ is the period of the external driving, $\lambda$ is the
quasi--energy, and
\be
A(I,\theta)=k\int_0^T V(I)g(\theta+\omega(I)t)
\cos(\Omega t) dt. \label{A}
\ee
(for details, see \cite{Brenner96}).
The physical
reason why the Floquet operator can be factored  into a product of two
operators, as in Eq. (\ref{kick}), is that most of the energy 
transfer takes place around the
singular point of the potential. When approximating a 3D potential
by one degree of freedom, usually one is dealing with the half line
so a special point naturally arises.
For the hydrogen atom, e.g., this
is the nucleus; this has been the basis for constructing the
Kepler map \cite{Gontis87,Casati88}. The existence of 
a singular point in space is
related to the slow decrease of the Fourier components of the driving,
$G(m)$. 
 
We define a dimensionless parameter 
$\epsilon\!=\!\omega/\Omega$, where $\omega\!=\!\del{\cal H}_0/\del I$.
In the absence of the driving, the solutions of 
Eq. (\ref{kick}) 
are $\delta$--functions in $n$-space.
For $\epsilon\ll 1$, 
the perturbation couples most effectively  $\delta$-functions
separated by approximately $j\hbar\Omega$ in energy 
($j$ integer), corresponding to nearly degenerate
$\lambda$'s. These form the QR ladder, which will be derived in what
follows.

For $\epsilon\ll 1$,
\be
\renewcommand{\arraystretch}{1.4}
A(I;\theta)=\left\{
\begin{array}{l}
0\hfill 0\!<\!\theta\!<\! 2\pi(1\!-\!\epsilon)\\
\cos\left[\frac{2\pi-\theta}{\epsilon}\right]~~~~~
\hfill 2\pi(1\!-\!\epsilon)\!<\!\theta\!<\!2\pi\\
\end{array}\right.  \label{Aapprox}
\ee
where $R=\frac{2\pi}{\omega} k V(I) G(m_+)$,
and $m_+$ is the integer closest to $1/\epsilon$, corresponding to
classical resonance $\Omega\!=\!m_+\omega$.
The Fourier series of $A$
is dominated by this resonance; the smoothness of the
driving coefficients allows us to take into account not only this term
but also its vicinity.

Since ${\cal H}_0(n\hbar)$ varies slowly as a
function of $n$, it can be expanded around a
large value $n_0$ to first order:
${\cal H}_0(n\hbar)\!\simeq\!{\cal H}_0(n_0\hbar)\!+\!\omega l\hbar$,
where $l\!\equiv\!(n\!-\!n_0)$.
Then the eigenvalue equation (\ref{kick}) 
appears as the equation for a ``linear kicked rotor''
\cite{lin1}, which is an exactly solvable model. Denoting
the quasi--energies of the linearized model by $\lambda^{lin}$,
its spectrum is $\lambda^{lin} T \equiv \mu \omega T (\mbox{mod}
 ~2\pi)$, and the corresponding eigenstates are
$|u\ket e^{-i\lambda^{lin}t}$, where
\be
\bra l|u\ket~~=~~e^{i\varphi}\sum_{m=-\infty}^{\infty}
J_m(B)~\mbox{sinc}[\pi(\mu-l+m/\epsilon)] \label{ul}
\ee
with the following definitions:
\renewcommand{\arraystretch}{1.4}
\be
\begin{array}{l}
\varphi=\pi(\mu-l)(1-2\epsilon)\\
B=\frac{\pi }{\hbar\omega}kV(n_o\hbar)
~ G(m_+)/ \sin(\pi/\epsilon)
\end{array}
\ee
and sinc$(x)=\sin(x)/x$.             
In the high-frequency limit, $\epsilon\ll 1$, the function
(\ref{ul}) is composed of a chain of peaks separated by
approximately the energy
of one photon, each weighted by an amplitude.
These quasi--resonances (QR)
are described by the function
\be
Q_j(l) = \frac{\sin[\pi(l\!-\!l_j\!+\!\delta_j)]}
{\pi(l\!-\!l_j\!+\!\delta_j)} \label{sinc}
\ee
where $l_j$ is the center of the QR and $\delta_j$ characterizes its
precise shape.
The $Q_j$'s constitute a nonorthogonal set which is probably complete,
however for a given quasi--energy, the corresponding eigenstate
(\ref{ul}) is composed only of a small fraction of the $Q_j$'s,
at positions determined by 
$\lambda^{lin}$ through the 
relation $E_{l_j}\!=\!\lambda^{lin}\hbar\!+\!j\hbar\Omega
\!-\!\hbar\omega\delta_j$, with $|\delta_j|\!<\!1/2$.
Assuming that at each point tails of the
sinc functions centered far--away  contribute incoherently, 
the absolute value
of the wave function is large on this ladder of QR's. 
The width of the QR is independent of the
driving field strength and of the density of
unperturbed states. This is in  contrast
to the width associated with the transition rate
given by Fermi's golden rule, and also in contrast
to the Rabi width, both of which are invalid approximations
in our regime of parameters.

The next step is to construct the eigenstates of the 
original nonlinear system by matching different 
energy regimes, in each of which a local solution holds.
In this construction,
there are two important scales in action space.
The first is the distance between classical
resonances: our local solution is valid only outside
the close vicinity of the classical resonances. 
The second is
the width of the resulting local solution, which is determined
by the Bessel function in Eq. (\ref{ul}). Once the linearized
local functions become wide enough to cover more than one
classical resonance, it is exptected that the different local
solution may interact and matching between the regions may
be employed. It turns out that for the two classes of models described 
below, this matching condition coincides with the Chirikov critetion
for resonance overlap \cite{Chirikov79}. This implies that for 
these cases, the matching
is valid inside classically chaotic connected regions of phase space.
Outside such regions, the decay of the eigenstates
is expected to be determined by classical bounds.

Under conditions that the matching is valid, we use 
\be
E_{n_j}\approx E_{n_0}+\left(\del E_n/\del n\right)_{n_0}l_j
\ee
to find that in the matched nonlinear solution, the QR appear
on the ladder of states $n_j$ satisfying
\be
E_{n_j}~=~\lambda\hbar+j\hbar\Omega
-\left(\del E_n/\del n\right)_{n_0}\delta_j. \label{nonlinqr}
\ee
Thus the eigenvalue $\lambda$ sets the origin of the
ladder for the corresponding eigenstate, and the QR's
are numbered by their position $j$ on this ladder. 
Each is described approximately by the function $Q_j(n)$
of Eq. (\ref{sinc}),
characterized by the peak position $n_j$ and the detuning
$|\delta_j|\!<\!1/2$,  both satisfying Eq. (\ref{nonlinqr}).
The matching 
may alter the amplitudes of the QR's,
therefore we write for the global eigenstate the following
approximate form:
\be
\bra n|u\ket=\sum_j A_j~
Q_j(n)~=\sum_j A_j~
\frac{\sin[\pi(n\!-\!n_j\!+\!\delta_j)]}
{\pi(n\!-\!n_j\!+\!\delta_j)}. \label{superposition}
\ee
This equation, together with (\ref{nonlinqr}) which defines the
parameters $n_j$ and $\delta_j$, constitute the main result
concerning the structure of the Floquet eigenstates on a small scale.
They predict, for a given 
$\lambda$,
the precise location and shape of the QR's of the corrresponding
eigenstate.

We now turn to determine the amplitudes $A_j$.
The Floquet eigenvalue equation  
can be equivalently
written in an extended Hilbert space of variables $(n,j)$
of which $|n,j\ket=|n\ket e^{-i j\Omega t}$ are basis functions
\cite{Shirley65}: 
\renewcommand{\arraystretch}{1.8}
\be
\begin{array}{l}
(E_n-\hbar j\Omega)\phi_{n,j}~+\\~~~~\frac{k}{2}
\sum_{n'} \bra n|\hat{{\cal O}}|n'\ket \left[\phi_{n',j+1}
+\phi_{n',j-1}\right]~=~\hbar \lambda \phi_{n,j}
\end{array}
\ee
where $\phi_{n,j}=\bra n,j|\phi\ket$ and ${\cal O}=V(I)g(\theta)$.
For smooth functions $V(I)$ the semiclassical matrix elements
may be written as $\bra n|\hat{{\cal O}}|n'\ket\approx V(I)G(n-n')$.
Using the local structure of the QR's and
the properties of the sinc functions on a lattice,
the equation reads: 
\be
\frac{2\hbar\omega}{\pi k V(\hbar n_j)}\sin(\pi\delta_j)A_j-
G^{+}A_{j+1} -G^{-}A_{j-1} \approx 0 \label{gen-ampeq}
\ee 
where $G^{\pm}\!=\!G(\frac{\mp 1}{\epsilon_{j\pm 1}}\!+\!\delta_{j\pm 1})$,
and $\epsilon_{j\pm 1}$ is
the value of $\epsilon$ at the $(j\pm 1)$-th QR. 
This is a tight--binding equation on the 1D lattice
of QR's labelled $j$. The diagonal potential is determined by the
$\delta_j$, which are the detunings of the energies  
$E_{n_j}$ from exact
resonance, and the hopping is related to 
matrix elements of the driving between neighboring
QR's.  The fact that there is only near--neighbor coupling is a result
of the driving being harmonic.

We consider two classes of models in some detail. These are
described by the Hamiltonian
\be
{\cal H}(p,x)=\frac{p^2}{2m}+b~x^{\sigma}+
k~\hat{\cal O}\cos(\Omega t),~~~~~~~~x\geq 0  \label{hamil2}
\ee
where $\sigma$ is a real number, $-2<\sigma<2$. The driving field
is described in the dipole approximation, and it is
convenient to work in different gauges for the cases of positive or
negative $\sigma$, thus $\hat{\cal O}\!=\!\hat{x}$ for $\sigma>0$,
and $\hat{\cal O}\!=\!\hat{p}/\Omega$ for $\sigma<0$.
Both classes have a slowly varying density of 
states, the interaction with the field is described by a function
$V(I)$ which is slowly varying with $I$, and $g(\theta)$ with smooth
Fourier coefficients. For high energies, (\ref{hamil2}) 
is well approximated by (\ref{hamil1}).
Thus, on the fine scale they exhibit the universal structure of the QR's. 
The differences between the models 
appear in the 
envelope of amplitudes superimposed on the peak structure.

For $\sigma>0$, the binding potentials have a triangle--like singularity
near the origin $x\!=\!0$. The special case of $\sigma\!=\!1$ corresponds
to the bubble model, which is a triangular potential well. 
The dipole matrix elements have the asymptotic form
$\bra n|\hat{x}|m\ket~\propto~(n\hbar)^{2/(2\!+\!\sigma)}/
(n-m)^2$.
The eigenstates for this case may be characterized by
two qualitatively different regimes in $n$-space. In the small $n$ regime
they are exponentially localized, while for larger values of
$n$ they are more extended, with a crossover point $n_c$ between the two
regions, satisfying
\be
n_c\sim (\Omega^2/k)^{(2+\sigma)/\sigma}.
\ee
This is a generalization of a result previously
found for the special case of the bubble model
\cite{Brenner95}.
In the asymptotic regime $n\to\infty$, Eq.
(\ref{gen-ampeq}) becomes
similar to the Anderson model in a static electric
field:
\renewcommand{\arraystretch}{1.9}
\be
\left(\frac{\sin(\pi\delta_j)}{\sqrt{j}}\right)
\left(\frac{\sqrt{\hbar}\Omega^{3/2}}{C_1k}\right)A_j
+ R^+ A_{j+1}+ R^- A_{j-1}\approx 0.
\label{oliv+}
\ee
where $C_1$ is a constant and $R^{\pm}=(1\mp 
\delta_{j\!\pm\!1}\epsilon_{j\!\pm\!1})^{-2}$.
The diagonal potential is given
by the pseudo--random function $\sin(\pi\delta_j)$, where
$\delta_j=\mbox{frac}\left\{ \frac{1}{\hbar}\left[
\hbar\alpha(\lambda+j)\right]
^{1/\alpha}\right\}$ and $\alpha\!=\!2\sigma/(2\!+\!\sigma)$,
damped along the lattice by $1/\sqrt{j}$.
The hopping terms depend weakly on position $j$. 
For a 
random potential and strictly constant hopping,
the eigenstates are power localized with a power
proportional to the square of the prefactor of the
diagonal potential \cite{Delyon84}.
Numerical calculations
for the model (\ref{oliv+}) with
$\sigma\!=\!1$, provide evidence that
these differences do not alter the qualitative conclusion
from the theorem. (\cite{Brenner95}, following \cite{Oliveira94}). Thus the
eigenstates are power--law decaying,
\be
A_j~\sim~1/j^{(\hbar\Omega^3/k^2)}.
\ee
This implies the existence of a crossover field strength
$k_c\sim \sqrt{\hbar\Omega^3}$ beyond which the tails
of the eigenstates turn non--normalizable.
A similar result was stated previously in \cite{Oliveira94}.
For the special case of the bubble model, it was
first obtained by use of a Kepler--like map \cite{Benvenuto91}.
The reason that in this regime the result is independent of
$\sigma$ is related to the dipole matrix elements being asymptotically
similar for all $\sigma$,
which is a result of the triangle--like singularity
that all these potentials have at the origin.

For $\sigma<0$, the potential is singular at the origin 
($\lim_{x\to 0}x^{\sigma}=-\infty$). The special case of
$\sigma=-1$ corresponds to the 1D Coulomb potential,
which is a good approximation for hydrogen--like rydberg atoms
with small angular momentum \cite{Casati87}. For this case, the
derivation of the universal QR structure is strictly justified only
as a leading order in the field strength $k$. Using the asymptotic form
$\bra n|\hat{p}|m\ket\propto i~  \mbox{sgn(n-m)}
(n\hbar)^{\sigma/2\!+\!\sigma}/
|n-m|^{2/2\!-\!\sigma}$, 
the resulting equation for the amplitudes is,
\be
\left(\frac{\hbar\Omega^{\eta}}{C_2 k}\right)
\sin(\pi\delta_r)
~A_r
~+~R^+ A_{r+1}
- R^- A_{r-1}~\approx~0
\label{holiv+}
\ee
where $C_2$ is a constant, $R^{\pm}=(1\mp \delta_{r\pm1}
\epsilon_{r\pm 1})^{\frac{2}{\sigma-2}}$, $\eta=(4-\sigma)/(2-\sigma)$
and 
$\delta_r=\mbox{frac}\left\{\frac{1}{\hbar}
\left[\alpha\left(\hbar\lambda \!-\!(r_{max}\!-\!r)\hbar\Omega
\right)\right]^{1/\alpha}\right\}$ for $r<r_{max}$; since the
bound part of the spectrum is of finite range in energy, there
is only a finite number of QR's. 
For $\epsilon<1$, the diagonal potential
may be considered
similar to a random one.  Neglecting the
weak dependence of the hopping terms on $r$, 
one obtains  a  1D
Anderson model for the amplitudes $A_r$.
The eigenstates of this model are
exponentially decaying with a perturbative estimate for
the localization length given by
\be
\xi \sim k^2/\hbar^2\Omega^{2\eta}. \label{hAnderson}
\ee
Also in this case, numerical calculations confirm that the
weak dependence of the hopping terms on position does not
alter the existence of exponential localization and the scaling
(\ref{hAnderson}). 
Our description
neglects completely effects of the continuum, and thus is valid
only far away from it; the result (\ref{hAnderson}) is meaningful
only if $\xi$ is smaller than the lattice size, which is finite in this case.
 
In conclusion, for a particle in a 1D
potential well, where the unperturbed spectrum is slowly varying, driven
by a harmonic high--frequency field, some
properties of the Floquet eigenstates were found on several
energy scales. The small scale
resonant structures, the QR's, are non--perturbative
and {\em universal}, independent of the details of the system and on the
parameters of the driving. Thus the general form of the eigenstate is
as described by Eqs. (\ref{nonlinqr}), (\ref{superposition}). 
On a larger scale, it was shown in general
how to construct a tight--binding equation, Eq. (\ref{gen-ampeq}),
for the amplitudes $A_j$ of (\ref{superposition}) superimposed on the QR's. 
Two specific classes were considered,
where this equation turns out to be an Anderson model
on a finite lattice, 
and an Anderson model on an infinite lattice with
a constant electric field.
It will be of great interest to study experimentally the nature of 
the eigenstates and in particular the shape of the 
quasiresonances for some systems that belong to classes 
explored in the present work. Recent developments in the 
optical trapping of atoms may enable such investigations 
for various potentials.

This research was supported in part by
the US-Israel Binational Science Foundation (BSF), 
by the National Science Foundation under Grant No. PHY94-07194,
by the Fund for Promotion of Research at the Technion, and by the E. and J.
Bishop Research Fund. 
We are grateful to O. Agam, I. Guarneri, B. Segev, D. Shepelyansky
and U. Smilansky for helpful discussions and comments.

\end{document}